\documentclass[referee]{raa}            
\usepackage{graphicx,times}             
\usepackage{natbib}
\usepackage{amssymb,amsmath}
\usepackage{datetime}
\usepackage{threeparttable}
\usepackage{multirow}

\bibpunct{(}{)}{;}{a}{}{,}

\usepackage[pagebackref=true]{hyperref}

\begin{document}

  \title{Pulsations of Three Rapidly Oscillating Ap Stars TIC~96315731, TIC~72392575, and TIC~318007796}

   \volnopage{Vol.0 (20xx) No.0, 000--000}      
   \setcounter{page}{1}          

\author{Hai-Jian Zhong  
  \inst{1}
  \and Dong-Xiang Shen
  \inst{1}
  \and Chun-Hua Zhu 
  \inst{1}
  \and He-Lei Liu
  \inst{1}
  \and Su-Fen Guo
  \inst{1}
  \and Guo-Liang L$\ddot{\rm u}$
  \inst{2, 1}
}

   \institute{School of Physical Science and Technology, Xinjiang University, 
              Urumqi 830046, China; {\it chunhuazhu@sina.cn}\\
        \and
        Xinjiang Astronomical Observatory, National Astronomical Observatories, Chinese Academy of Sciences, 
        Urumqi 830000, China; {\it guolianglv@xao.ac.cn}\\
\vs\no
   {\small Received 20xx month day; accepted 20xx month day}}

\abstract{We analyze the frequencies of three known roAp stars, TIC~96315731, TIC~72392575, and TIC~318007796, using the light curves from the Transiting Exoplanet Survey Satellite (TESS). For TIC~96315731, the rotational and pulsational frequencies are $0.1498360\,\mathrm{d}^{-1}$ and $165.2609\,\mathrm{d}^{-1}$, respectively. In the case of TIC~72392575, the rotational frequency is $0.25551\,\mathrm{d}^{-1}$. We detect a quintuplet of pulsation frequencies with a center frequency of $135.9233\,\mathrm{d}^{-1}$, along with two signals within the second pair of rotational sidelobes of the quintuplet separated by the rotation frequency. These two signals may correspond to the frequencies of a dipole mode. In TIC~318007796, the rotational and pulsational frequencies are $0.2475021\,\mathrm{d}^{-1}$, $192.73995\,\mathrm{d}^{-1}$, and $196.33065\,\mathrm{d}^{-1}$, respectively. Based on the oblique pulsator model, we calculate the rotation inclination $\left( i \right)$ and magnetic obliquity angle $\left( \beta \right)$ for the stars, which provides the geometry of the pulsation modes. Combining the phases of the frequency quintuplets, the pulsation amplitude and phase modulation curves, and the results of spherical harmonic decomposition, we conclude that the pulsation modes of frequency quintuplets in TIC~96315731, TIC~72392575, and TIC~318007796 correspond to distorted dipole mode, distorted quadrupole mode, and distorted dipole mode, respectively.
}

   \authorrunning{H.-J. Zhong, D.-X. Shen, C.-H. Zhu, H.-L. Liu, S.-F. Guo \& G.-L. L$\ddot{\rm u}$} 
   \titlerunning{Pulsations of Three Rapidly Oscillating Ap Stars}  

   \maketitle

%
%
\section{Introduction}\label{sect:intro}           

Chemically peculiar stars with spectral type A, known as Ap stars, typically exhibit magnetic field with intensities ranging from 300 G to 34 kG \citep{1960ApJ+Babcock, 2007A&A+Auri,2010MNRAS+Elkin}. The intense magnetic field suppresses convection in stellar interior, which leads to an interaction of gravitational settling and radiative levitation in the atmosphere of Ap stars \citep{1970ApJ+Michaud}. The intricate interaction results in their surfaces manifesting an overabundance of elements such as Pr, Nd, Eu, Gd, and Tb, which can reach values up to a million times that of the solar value \citep{2004A&A+Ryabchikova,2015ads+Michaud}. The overabundances of rare earth elements form chemical spots near their magnetic poles \citep{2007A&A+473+907R}, which remain stable over periods ranging from days to decades \citep{2009A&A+Renson}. From the light curves modulated by the spots, we can determine the rotation period of the stars.

Rapidly oscillating Ap (roAp) stars are a rare subset within the Ap stars \citep{1982MNRAS+200+807K}, located at the lower boundary of the classical instability strip and intersecting the main sequence. These stars exhibit high-overtone $\left( n\geqslant 15\right)$, low-degree $\left(l\leqslant 3\right)$ non-radial pressure mode (p-mode) pulsations with periods ranging from 4.7 to 25.8 minutes \citep[e.g.][]{2019MNRAS+Cunha, 2021MNRAS+506+1073H}. In the Johnson \emph{B} band, the semi-amplitudes of these pulsations can reach up to 18 mmag \citep{2015PhDT+227H}. In the majority of roAp stars, pulsations are believed to be driven by the $\kappa$-mechanism within the H \uppercase\expandafter{\romannumeral1} ionization zone \citep[e.g.][]{1996ApJ+458+338D, 1998MNRAS+301+31G, 2001MNRAS+323+362B, 2002MNRAS+333+47C}. However, for some roAp stars with pulsation frequencies exceeding the theoretical acoustic cut-off frequency limit, their pulsations are thought to be driven by turbulent pressure in the envelope region where convection is not suppressed \citep{2013MNRAS+436+1639C}. 

\citet{1982MNRAS+200+807K} introduced the oblique pulsator model for roAp stars, which was further developed by incorporating the influence of both rotation and magnetic fields \citep[e.g.][]{1985ApJ+296L+27D, 1993PASJ+45+617S, 1994PASJ+46+301T, 1995PASJ+47+219T, 2002A&A+391+235B, 2004MNRAS+350+485S, 2011A&A+536A+73B}. In this model, the pulsation axis is misaligned with the rotation axis, typically aligning with the magnetic axis. As a roAp star rotates (with a non-zero rotation inclination angle), the observed pulsation amplitude and phase undergo modulation. This modulation causes the pulsation frequency to split into $2l+1$ frequency components, each separated by an interval equivalent to the rotation frequency \citep{1982MNRAS+200+807K}. This not only provides geometric information about the pulsation modes but also facilitates pattern recognition. This peculiarity of roAp stars makes them excellent subjects for studying the interaction between magnetic fields and stellar pulsations.

After \citet{1978IBVS+1436+1K} discovered the first roAp star, the search for roAp stars has been ongoing \citep[e.g.][]{1991MNRAS+250+666M, 2011MNRAS+410+517B, 2014MNRAS+439+2078H, 2019MNRAS+488+18H, 2021FrASS+8+31H, 2024MNRAS+527+9548H}. In these search efforts, ground-based telescopes face some disadvantageous aspects in detecting low-brightness and low-amplitude roAp stars due to Earth's rotation and atmospheric effects. However, space telescopes such as Kepler \citep{2010ApJ+713L+79K} and the Transiting Exoplanet Survey Satellite \citep[TESS;][]{2015JATIS...1a4003R} with continuous and high-precision observation capabilities, can effectively overcome these disadvantages. Especially TESS, with its all-sky coverage and high-precision short-cadence photometric observation capabilities, has propelled the exploration of roAp stars. The latest observations from TESS indicate that the confirmed count of roAp stars has now reached 118 \citep{2024MNRAS+527+9548H}. Considering the extremely limited number of known roAp stars, a reevaluation of the pulsations exhibited by these acknowledged roAp stars using TESS's latest observational data will aid in revealing potential pulsation signals that may exist but have yet to be detected, thus enabling a more comprehensive understanding of the pulsation phenomena associated with roAp stars.

In this work, we report the results of frequency analysis, pulsation mode identification, and the geometry of pulsation modes for three roAp stars: TIC~96315731, TIC~72392575, and TIC~318007796. In Section~\ref{sect:Obs}, we introduce the three identified roAp stars, along with the TESS observations. A detailed analysis of frequency characteristics is provided in Section~\ref{sect:FreAnalysis}, while Section~\ref{PulMGeo} presents pulsation modes and their geometry. The characteristics of pulsation amplitude and phase modulation curves are discussed in Section~\ref{PulAmpandPhaMod}, and Section~\ref{SHD} presents the spherical harmonic decomposition results of the amplitude spectra. Finally, we present the conclusions for the stars in Section~\ref{DisandConclus}.

\section{The roAp Stars and TESS Observations}\label{sect:Obs}

\subsection{Three Known roAp Stars}

TIC~96315731 (HD~51203) is classified as an $\alpha^2\mathrm{CVn}$ star with a spectral type of A2p SrEuCr \citep{1982mcts+book+H, 1991A&AS+89+429R}. According to the report by \citet{2018MNRAS+478+4513B}, the effective temperature of the star is $7098 \pm 105\,\mathrm{K}$. \citet{2021MNRAS+506+1073H} identified distinct rotational and pulsational features in the light curves from TESS sectors 6 and 7, officially designating the star as a new roAp star. They determined the stellar rotation period as $6.6713\pm 0.0007\,\mathrm{d}$, consistent with the value reported by 
\citet{2015A&A+581A+138B} as $6.6752\pm 0.0008\,\mathrm{d}$. Additionally, \citet{2021MNRAS+506+1073H} determined the frequency of the central peak in the frequency quintuplet is $165.258\pm 0.001\,\mathrm{d}^{-1}$ $\left(1.91271\pm 0.00001\,\mathrm{mHz}\right) $. \citet{2019ApJ+873L+5C} reported a mean magnetic field modulus of $7.9\pm 0.5\,\mathrm{kG}$ for TIC~96315731 by analyzing magnetically split absorption lines in the H-band spectra obtained from the Apache Point Observatory Galactic Evolution Experiment (APOGEE) survey.

TIC~72392575 (HD~225578) exhibits an ambiguity in its spectral classification. Initially, it was classified as ApSr: by \citet{1964PASP+76+119C}, but later~\citet{1991A&AS+89+429R} determined it as A5 Sr\@. \citet{2012A&A+542A+89P} reported the effective temperature of the star to be $8290\pm 60\,\mathrm{K}$\@. \citet{2012A&A+542A+89P} used photometric data from the 1 m Austrian-Croatian Telescope (ACT) with a Bessell \emph{B} filter system over 130.9 minutes and did not detect any pulsation signals, with the corresponding maximum amplitude in the Fourier spectrum was found to be 1.6 mmag. Based on the TESS light curves, \citet{2022arXiv221210776B} initially identified this star as a roAp star and derived its rotation period as 3.922 d. Recently, \citet{2024MNRAS+527+9548H} detected a pulsation signal in the TESS sector 14 light curve at a frequency of $135.944\pm 0.003 \,\mathrm{d}$ $\left(1.57343\pm 0.00003 \,\mathrm{mHz}\right) $, with no evidence of pulsation frequency rotational sidelobes. They also determined the rotation period of the star as $3.9016\pm 0.0008 \,\mathrm{d}$. 

TIC~318007796 (HD~190290) is classified as an $\alpha ^2\mathrm{CVn}$ star with a spectral type of A2 EuSrCr \citep{1993IBVS+3840+1K, 1999A&A+351+283P}. The Strömgren-Crawford indices measured by \citet{2016A&A...590A.116J} for the star are as follows: $b-y = 0.289$, $m_1 = 0.293$, $c_1 = 0.466$ and $\mathrm{H}\beta =2.796$, giving $\delta m_1=-0.091$ and $\delta c_1=-0.306$. According to the report by \citet{2023ApJ+955+123S}, the effective temperature of the star is $7079 \pm 163\, \mathrm{K}$\@. \citet{1990IBVS+3510+1M} obtained high-speed photometric data for 8.85 hours at the South African Astronomical Observatory (SAAO) by the 1.0-m Elizabeth telescope with a Johnson \emph{B} filter. They identified rapid oscillations at frequencies $196.13\,\mathrm{d^{-1}}\left(2.27 \,\mathrm{mHz}\right)$ and $192.67\,\mathrm{d^{-1}}\left(2.23 \,\mathrm{mHz}\right)$, revealing a frequency separation of $40 \,\mathrm{\mu Hz}$ between the two signals. Furthermore, after pre-whitening these frequencies, the amplitude spectrum hinted at additional unresolved frequencies\@. \citet{2012MNRAS+426+969V} conducted differential photometric observations and derived a rotation period of $4.03\pm 0.13 \,\mathrm{d}$ but did not mention pulsation frequencies. Recently, \citet{2024MNRAS+527+9548H} detected pulsation signals with frequencies of $192.73997\pm 0.00013\,\mathrm{d}$ $\left(2.2307867\pm 0.0000015\,\mathrm{mHz}\right)$ and $196.33370\pm 0.00002\,\mathrm{d}$ $\left(2.2723808\pm 0.0000002 \,\mathrm{mHz}\right)$ in TIC~318007796, using light curves from TESS sectors 27 and 39. The rotation period of the star was determined to be $4.09070\pm 0.00002 \,\mathrm{d}$. They emphasized that the mode with the lower frequency corresponds to a distorted dipole mode, while the mode at the other frequency represents a distorted quadrupole mode. This suggests a large frequency separation of approximately $83 \,\mathrm{\mu Hz}$. Furthermore, \citet{2004A&A+415+685H} and \citet{2006A&A+450+777B, 2015A&A+583A+115B}conducted measurements on the mean longitudinal magnetic field of TIC~318007796. Based on these measurements, \citet{2023RAA+23i5024R} reported an averaged quadratic effective magnetic field of $\left< B_{\mathrm{e}} \right> = 3104\pm 182\,\mathrm{G}$.

\begin{table}[t]
  \centering
  \begin{threeparttable}
    \caption[]{Detailed Information of TESS Light Curves for TIC~96315731, TIC~72392575, and TIC~318007796}\label{Tab:1}
    \begin{tabular}{lccccc}
      \hline\noalign{\smallskip}
      Name          & Sector & BJD                           & Time span & Number of points & Exposure time \\
                    &        &                               & (d)       &                  & (s)           \\
      \hline\noalign{\smallskip}
      TIC~96315731  & 6      & 2458468.27520~-~2458490.04775 & 21.77     & 14639            & 120           \\
      TIC~96315731  & 7      & 2458491.63525~-~2458516.08916 & 24.45     & 16084            & 120           \\
      TIC~96315731  & 33     & 2459201.73606~-~2459227.57521 & 25.84     & 17458            & 120           \\
      TIC~96315731  & 34     & 2459229.19326~-~2459254.06781 & 24.87     & 16689            & 120           \\
      TIC~72392575  & 54     & 2459769.90234~-~2459796.04423 & 26.14     & 17417            & 120           \\
      TIC~72392575  & 55     & 2459797.10394~-~2459824.26598 & 27.16     & 18882            & 120           \\
      TIC~318007796 & 27     & 2459036.28194~-~2459060.64411 & 24.36     & 16781            & 120           \\
      TIC~318007796 & 39     & 2459361.77426~-~2459389.72065 & 27.95     & 19337            & 120           \\
      \noalign{\smallskip}\hline
    \end{tabular}{\footnotesize \textbf{Note.} The first column gives the stellar names, the second column lists the light curve-associated sectors, and the third to fifth columns provide the start and end times of the light curves, the duration, and the count of points, respectively, with the sixth column specifying the exposure time.}
  \end{threeparttable}
\end{table}

\subsection{TESS Observations and Data Reduction}

Our targets are observed in different sectors by TESS\@. In detail, TIC~96315731 is observed in sectors 6, 7, 33, and 34; TIC~72392575 in sectors 54 and 55; and TIC~318007796 in sectors 27 and 39. The detailed information of the light curves for these sectors is shown in Table~\ref{Tab:1}.  We employ the Python package \emph{Lightkurve} \citep{2018ascl.soft12013L} to download 2-minute cadence light curves of the stars processed by the Science Processing Operations Center (SPOC) from the Mikulski Archive for Space Telescopes (MAST) server. 

\begin{figure}[!t]
  \flushright{}
  \begin{subfigure}[t]{\textwidth}
    \includegraphics[width=\linewidth]{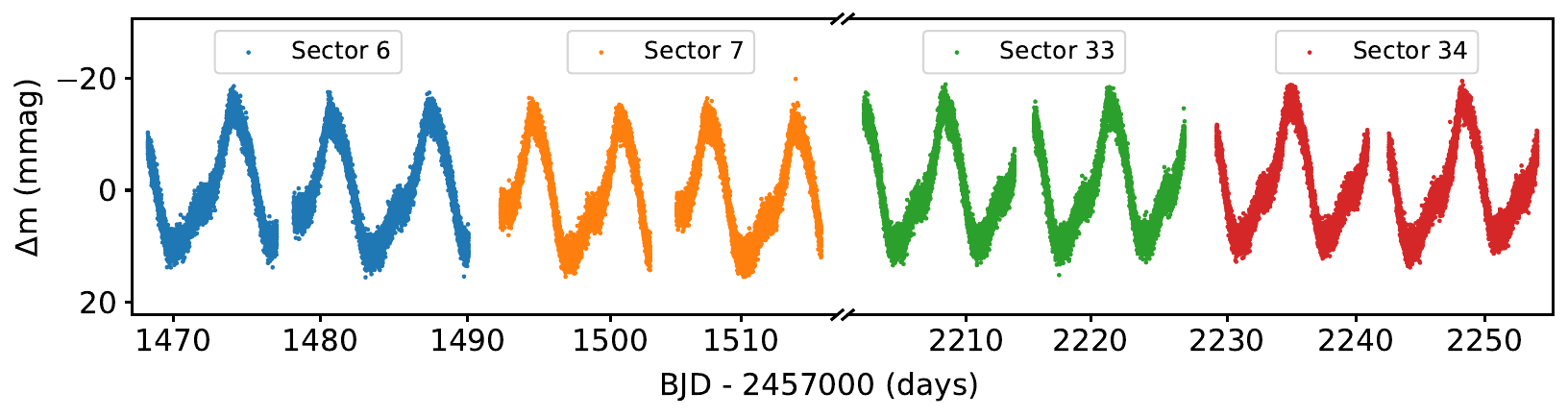}
  \end{subfigure}
  \hfill
  \begin{subfigure}[t]{\textwidth}
    \includegraphics[width=\linewidth]{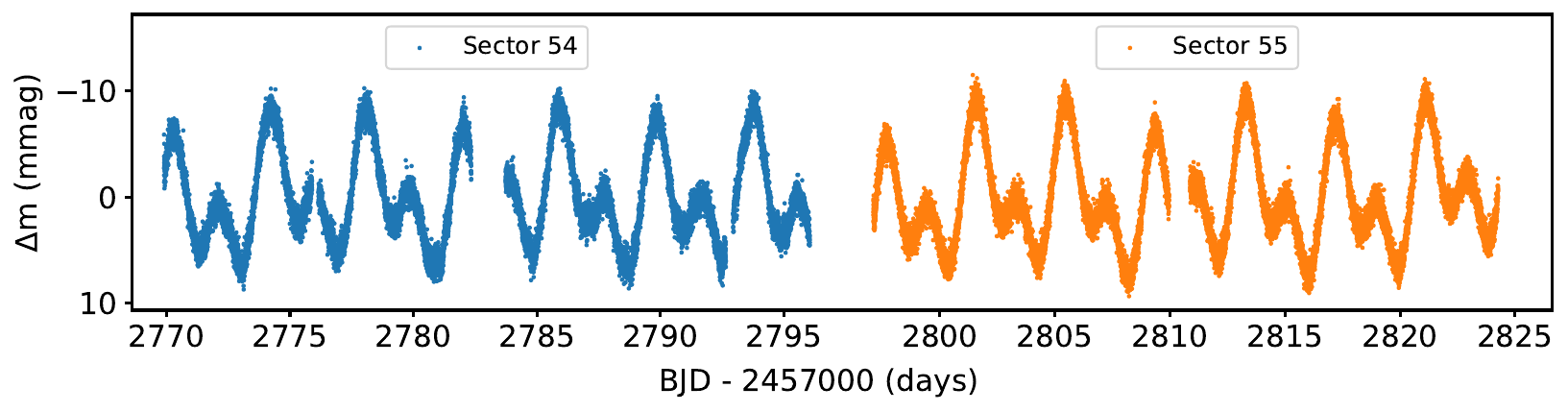}
  \end{subfigure}
  \hfill
  \begin{subfigure}[t]{\textwidth}
    \includegraphics[width=\linewidth]{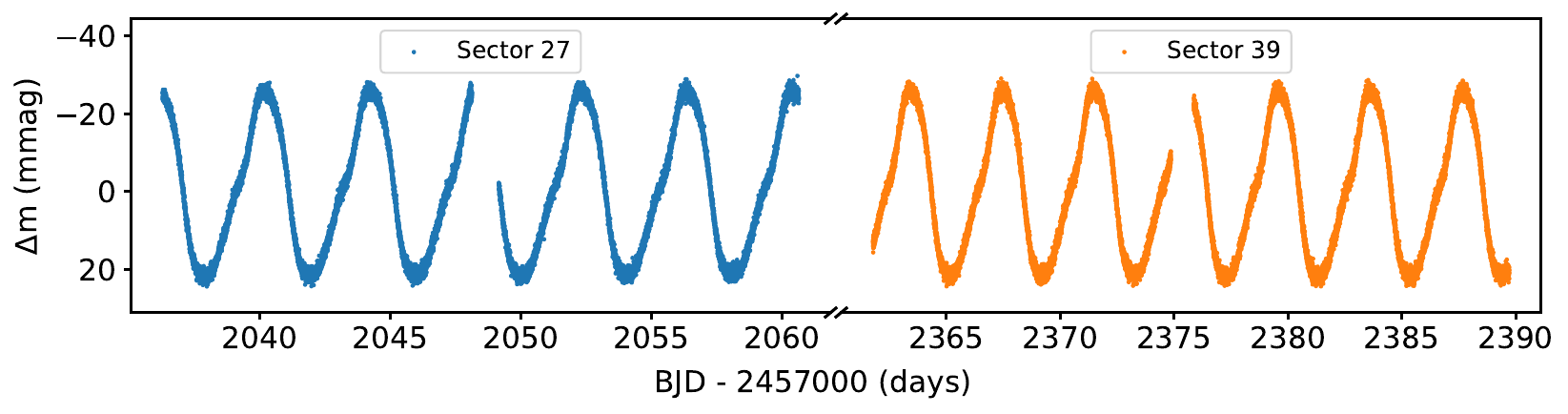}
  \end{subfigure}
  \caption{Top panel: Light curves of TIC~96315731 with four different sectors; Middle panel: Light curves of TIC~72392575 with two different sectors; Bottom panel: Light curves of TIC~318007796 with two different sectors.}\label{Figure:1}
\end{figure}

The PDC\_SAP fluxes of the light curves are used in our analysis. The \emph{Lightkurve} package is also used to remove outliers greater than $5\sigma$ from the light curves, as well as other points with quality issues, such as sudden changes in background measurements and impulsive outliers. The light curves for each sector are then divided by their median flux to achieve flux normalization, and the fluxes are subsequently converted to relative magnitudes. For TIC~96315731, we combine the light curves of sectors 6, 7, 33, and 34, resulting in a total of 64780 data points over an observational period of 96.94 days. Similarly, for TIC~72392575, sectors 54 and 55 are combined, yielding 36299 data points observed over 53.30 days. For TIC~318007796, sectors 27 and 39 are combined, comprising 36118 data points observed over 52.31 days. Fig.~\ref{Figure:1} displays the combined light curves of the stars. It is evident that the combined light curves of the stars display distinct rotational features, which are controlled by the spots on their surfaces. 

\section{Frequency Analysis}\label{sect:FreAnalysis}

We perform a Discrete Fourier Transform \citep[DFT;][]{1985MNRAS+213+773K} on the combined light curves of TIC~96315731, TIC~72392575, and TIC~318007796 to obtain their rotation and pulsation frequency information. The Nyquist frequency $f_{\mathrm{Nyq}}=1/(2\Delta t)$, where $\Delta t$ is the sampling time interval of TESS observations, is the highest frequency without undersampling at the given sampling rate. For TESS observations, the Nyquist frequency is approximately $359\,\mathrm{d}^{-1}$. Consequently, the frequency range of the amplitude spectrum spans is $0-359\,\mathrm{d}^{-1}$. When the frequency separation between two frequencies exceeds the frequency resolution, defined as $f_{\mathrm{res}} = {1.5}/{T}$, they become distinguishable \citep{1978Ap&SS+56+285L}. Here, $T$ is the time span of the combined light curve. For TIC~96315731, TIC~72392575, and TIC~318007796, their respective frequency resolutions are $0.01547\,\mathrm{d}^{-1}$, $0.0281\,\mathrm{d}^{-1}$, and $0.02868\,\mathrm{d}^{-1}$.

\begin{figure}[htbp]
  \centering
  \includegraphics[width=\textwidth, angle=0]{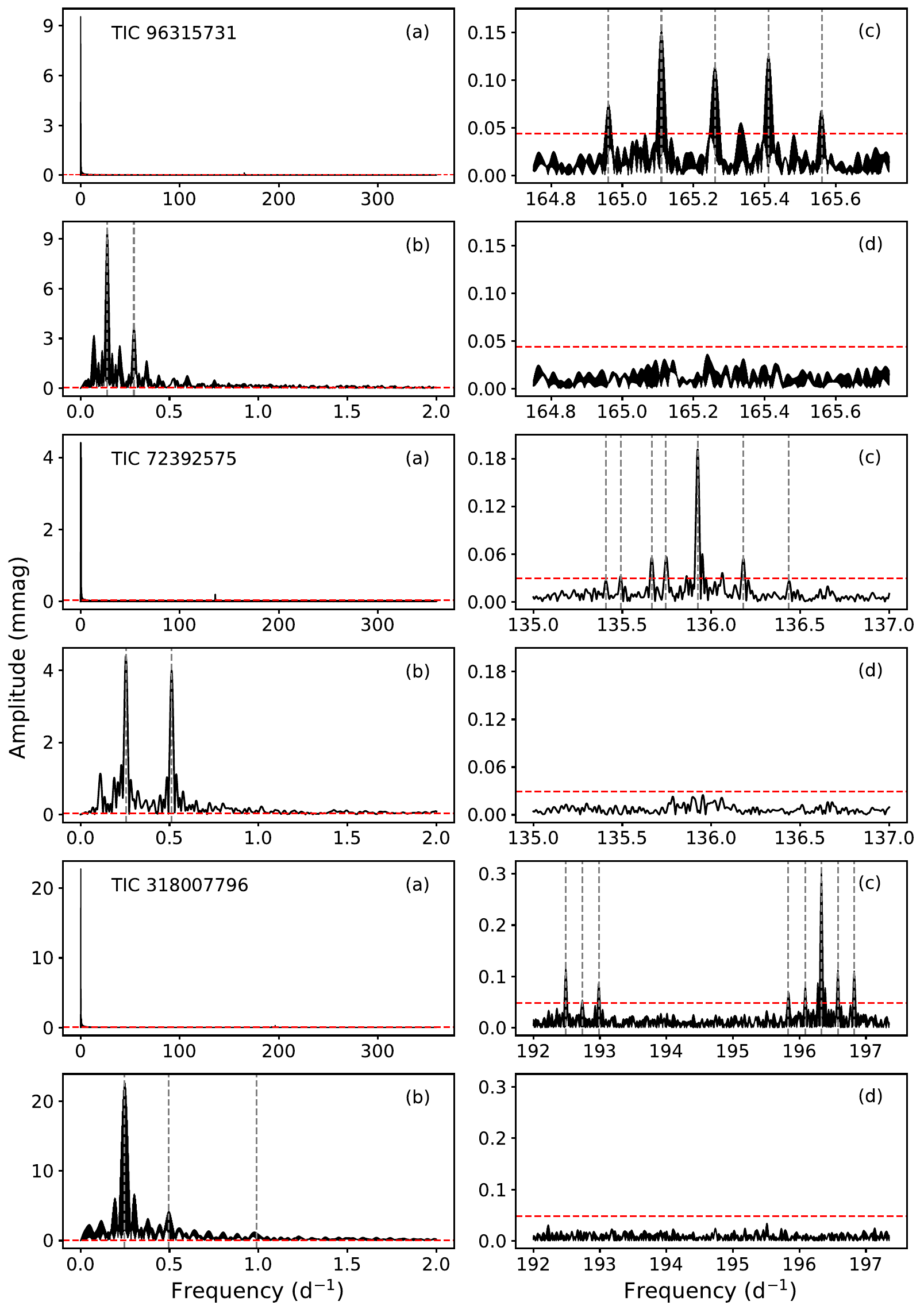}
  \caption{The amplitude spectra of TIC~96315731, TIC~72392575, and TIC~318007796.~(a)~The amplitude spectra within the frequency range from $0-359\,\mathrm{d}^{-1}$.~(b)~Amplitude spectra within the frequency range of $0-2\,\mathrm{d}^{-1}$.~(c)~Amplitude spectra of the region containing pulsation frequency after pre-whitening of low-frequency signals.~(d)~The residual amplitude spectra after pre-whitening the pulsation frequency. The red dashed lines indicate 4.6 times the white noise, while the gray dashed lines denote the extracted frequencies.
  }\label{Figure:2}
\end{figure}

Fig.~\ref{Figure:2} illustrates the amplitude spectra of TIC~96315731, TIC~72392575, and TIC~318007796. Panels (a) display the amplitude spectra covering the frequency range of $0-359\,\mathrm{d}^{-1}$. Panels (b) exhibit signals related to the rotational frequency, while panels (c) show signals associated with the pulsation frequency. The residual amplitude spectra after pre-whitening the pulsation frequency are presented in panels (d). The red dashed lines represent 4.6 times the background signal, while the gray dashed lines denote the extracted frequencies.

\subsection{Rotation Frequency Analysis}\label{sect:RotFreAnalysis}

During the analysis of the rotation frequency, we apply the DFT to extract both the rotation frequency and its harmonics within the frequency range of $0-2\,\mathrm{d}^{-1}$. Following this, linear least squares fitting is employed to determine the amplitude and phase values associated with the identified frequencies. Lastly, optimization of the frequency, amplitude, and phase results is conducted using nonlinear least squares fitting technique \citep[same with the technique in][]{1982MNRAS+200+807K}. The results of the nonlinear least squares fitting for TIC~96315731, TIC~72392575, and TIC~318007796 are presented in Table~\ref{Tab:2}. 

\subsubsection{TIC~96315731}\label{sect:RotFreAnalysis5731}

For TIC~96315731, the amplitude spectrum in the low-frequency region is shown in Fig.~\ref{Figure:2} (b).The results of the rotation frequency and its first harmonic are presented in Table~\ref{Tab:2}. The derived rotation frequency of the star is $\nu_{\mathrm{rot}}=0.1498360\pm 0.0000005\,\mathrm{d}^{-1}$ $\left( P_{\mathrm{rot}}=6.67396\pm 0.00002\,\mathrm{d} \right)$. It is noteworthy that the formal error reported here is calculated using the method proposed by \citet{1999DSSN+13+28M}, which represents the lower limit of the error. Our rotation period is consistent with the findings reported by \citet{2015A&A+581A+138B}, who measured a rotation period of $6.6752\pm 0.0008\,\mathrm{d}$, and by \citet{2021MNRAS+506+1073H}, who recorded a rotation period of $6.6713\pm 0.0007\,\mathrm{d}$.

\subsubsection{TIC~72392575}\label{sect:RotFreAnalysis2575}

For TIC~72392575, the amplitude spectrum in the low-frequency region is depicted in Fig.~\ref{Figure:2}~(b). Table~\ref{Tab:2} presents the results of the rotation frequency and its first harmonic. The derived rotation frequency of the star is $\nu_{\mathrm{rot}}=0.25551\pm 0.00002\,\mathrm{d}^{-1}$ $\left(P_{\mathrm{rot}}=3.9138\pm 0.0004\,\mathrm{d}\right)$, Our result is consistent with the rotation period of $3.9016\pm 0.0008\,\mathrm{d}$ determined by \citet{2024MNRAS+527+9548H}. Furthermore, our analysis benefits from a longer overall observational period, which results in reducing uncertainty in the derived rotational period.

\subsubsection{TIC~318007796}\label{sect:RotFreAnalysis7796}

For TIC~318007796, Fig.~\ref{Figure:2}~(b) displays the amplitude spectrum in the low-frequency region.The results of the rotation frequency and its first harmonic are shown in Table~\ref{Tab:2}. The derived rotation frequency of $\nu_{\mathrm{rot}}=0.2475021\pm 0.0000004\,\mathrm{d}^{-1}$ $\left(P_{\mathrm{rot}}=4.040370\pm 0.000007\,\mathrm{d}\right)$. Within the uncertainties, our rotation period is consistent with the one derived by \citet{2012MNRAS+426+969V}, which is $4.03\pm 0.13\,\mathrm{d}$. However, it is smaller than the rotation period of $4.09070\pm 0.00002\,\mathrm{d}$ obtained by \citet{2024MNRAS+527+9548H}. This discrepancy may mainly come from the different research methods.  

\subsection{Pulsation Frequency Analysis}\label{PulFreAnalysis}

Before determining the pulsation frequencies, we perform high-pass filtering on the combined light curves of the stars to remove the effects of low-frequency signals such as stellar rotation and instrumental long-term trends. This ensures the accuracy of analyzing high-frequency pulsation signals. High-pass filtering is achieved by continuously pre-whitening peaks within the frequency range of $0-20\,\mathrm{d}^{-1}$ until their amplitude levels are comparable to the noise level in the high-frequency region. Prior to high-pass filtering, careful examination is conducted to ascertain the absence of astrophysical signals within the high-frequency region. We utilize the batch mode of the Period04 software \citep{2005CoAst+146+53L} to execute the high-pass filtering. Similar to the extraction of rotation frequencies, we perform DFT on the high-pass filtered combined light curves to obtain pulsation frequencies for each star. Amplitude and phase values of the pulsation frequencies are then obtained through linear least squares fitting, followed by the application of nonlinear least squares fitting to optimize the frequencies, amplitudes, and phases \citep[same with the technique in][]{1982MNRAS+200+807K}.

\begin{table}[!b]
  \centering
  \begin{threeparttable}
    \caption[]{Nonlinear Least Squares Fitting Results for All Extracted Frequencies in TIC~96315731, TIC~72392575, and TIC~318007796
    }\label{Tab:2}
    \begin{tabular}{clccc}
      \hline\noalign{\smallskip}
      Star                            & Label                          & Frequency                      & Amplitude                    & Phase                       \\
                                      &                                & $\left(\mathrm{d}^{-1}\right)$ & $\left(\mathrm{mmag}\right)$ & $\left(\mathrm{rad}\right)$ \\
      \hline\noalign{\smallskip}
      \multirow{7}{*}{TIC~96315731}   & $\nu _{\mathrm{rot}}$          & 0.1498360 $\pm$ 0.0000005      & 10.40$\pm$0.01               & 3.154 $\pm$ 0.001           \\
                                      & $2\nu _{\mathrm{rot}}$         & 0.2996231 $\pm$ 0.0000014      & 3.43$\pm$0.01                & 2.190 $\pm$ 0.003           \\

                                      & $\nu-2\nu _{\mathrm{rot}}$     & 164.95976 $\pm$ 0.00005        & 0.063$\pm$0.007              & -1.26$\pm$0.12              \\
                                      & $\nu-\nu _{\mathrm{rot}}$      & 165.10972 $\pm$ 0.00002        & 0.161$\pm$0.007              & 1.98$\pm$0.05               \\
                                      & $\nu$                          & 165.26090 $\pm$ 0.00003        & 0.108$\pm$0.008              & -1.93$\pm$0.07              \\
                                      & $\nu+\nu _{\mathrm{rot}}$      & 165.41075 $\pm$ 0.00002        & 0.139$\pm$0.007              & -1.15$\pm$0.05              \\
                                      & $\nu+2\nu _{\mathrm{rot}}$     & 165.56060 $\pm$ 0.00005        & 0.061$\pm$0.007              & 2.69$\pm$0.13               \\
      \noalign{\smallskip}\hline\noalign{\smallskip}
      \multirow{9}{*}{TIC~72392575}   & $ \nu_{\mathrm{rot}}         $ & 0.25551  $\pm$ 0.00002         & 4.19$\pm$0.01                & 2.289 $\pm$ 0.002           \\
                                      & $2\nu_{\mathrm{rot}}         $ & 0.51166  $\pm$ 0.00002         & 3.91$\pm$0.01                & 2.277 $\pm$ 0.003           \\

                                      & $\nu_{1}                     $ & 135.4916 $\pm$ 0.0023          & 0.022$\pm$0.005              & -1.41 $\pm$ 0.23            \\
                                      & $\nu_{2}                     $ & 135.7454 $\pm$ 0.0009          & 0.056$\pm$0.005              & -0.90 $\pm$ 0.09            \\

                                      & $\nu_{3}-2\nu _{\mathrm{rot}}$ & 135.4089 $\pm$ 0.0019          & 0.026$\pm$0.005              & 2.36  $\pm$ 0.19            \\
                                      & $\nu_{3}- \nu _{\mathrm{rot}}$ & 135.6659 $\pm$ 0.0010          & 0.053$\pm$0.005              & 2.41  $\pm$ 0.10            \\
                                      & $\nu_{3}                     $ & 135.9233 $\pm$ 0.0003          & 0.190$\pm$0.005              & 2.60  $\pm$ 0.03            \\
                                      & $\nu_{3}+ \nu _{\mathrm{rot}}$ & 136.1785 $\pm$ 0.0010          & 0.052$\pm$0.005              & 2.41  $\pm$ 0.10            \\
                                      & $\nu_{3}+2\nu _{\mathrm{rot}}$ & 136.4358 $\pm$ 0.0024          & 0.021$\pm$0.005              & 2.41  $\pm$ 0.24            \\
      \noalign{\smallskip}\hline\noalign{\smallskip}
      \multirow{11}{*}{TIC~318007796} & $\nu _{\mathrm{rot}}         $ & 0.2475021 $\pm$ 0.0000004      & 23.208$\pm$0.009             & 2.7989 $\pm$ 0.0004         \\
                                      & $2\nu _{\mathrm{rot}}        $ & 0.4950086 $\pm$ 0.0000020      & 4.282$\pm$0.009              & 1.2453 $\pm$ 0.0020         \\
                                      & $4\nu _{\mathrm{rot}}        $ & 0.9900158 $\pm$ 0.0000070      & 1.219$\pm$0.009              & 2.9984 $\pm$ 0.0072         \\

                                      & $\nu_{1}-\nu _{\mathrm{rot}} $ & 192.48973 $\pm$ 0.00006        & 0.115$\pm$0.007              & 1.21   $\pm$ 0.06           \\
                                      & $\nu_{1}                     $ & 192.73995 $\pm$ 0.00013        & 0.050$\pm$0.007              & -0.94  $\pm$ 0.14           \\
                                      & $\nu_{1}+\nu _{\mathrm{rot}} $ & 192.98758 $\pm$ 0.00007        & 0.094$\pm$0.007              & -1.65  $\pm$ 0.07           \\

                                      & $\nu_{2}-2\nu _{\mathrm{rot}}$ & 195.83561 $\pm$ 0.00011        & 0.061$\pm$0.007              & 2.66   $\pm$ 0.11           \\
                                      & $\nu_{2}-\nu _{\mathrm{rot}} $ & 196.08629 $\pm$ 0.00008        & 0.085$\pm$0.007              & 1.33   $\pm$ 0.08           \\
                                      & $\nu_{2}                     $ & 196.33065 $\pm$ 0.00002        & 0.310$\pm$0.007              & -3.07  $\pm$ 0.02           \\
                                      & $\nu_{2}+\nu _{\mathrm{rot}} $ & 196.57818 $\pm$ 0.00006        & 0.106$\pm$0.007              & -1.81  $\pm$ 0.06           \\
                                      & $\nu_{2}+2\nu _{\mathrm{rot}}$ & 196.82260 $\pm$ 0.00006        & 0.111$\pm$0.007              & 0.33   $\pm$ 0.06           \\
      \noalign{\smallskip}\hline
    \end{tabular}{\footnotesize \textbf{Note.} The time zero-points for fitting to TIC~96315731, TIC~72392575, and TIC~318007796 are BJD~2,458,861.12908, BJD~2,459,797.32875, and BJD~2,459,213.62661, respectively. The table columns list star names, frequency labels, frequencies, amplitudes, and phases, in that order.}
  \end{threeparttable}
\end{table}

\subsubsection{TIC~96315731}\label{sect:PulFreAnalysis5731}

The amplitude spectrum after high-pass filtering, as depicted in Fig.~\ref{Figure:2} (c), reveals that TIC~96315731 exhibits a quintuplet of pulsation frequencies centered at $\nu =165.2609\,\mathrm{d}^{-1}$. Additionally, rotational sidelobes are observed at $\nu \pm \nu _{\mathrm{rot}}$ and $\nu \pm 2\nu _{\mathrm{rot}}$. The results for pulsation frequencies are presented in Table~\ref{Tab:2}. After pre-whitening the quintuplet of frequencies, a slight power excess is observed in the residual amplitude spectrum, as shown in Fig.~\ref{Figure:2} (d). The power excess suggests the presence of unresolved signals in the amplitude spectrum. 

\subsubsection{TIC~72392575}\label{sect:PulFreAnalysis2575}

From Fig.~\ref{Figure:2}~(c), the amplitude spectrum of TIC~72392575 after high-pass filtering indicates the presence of three pulsation frequencies: $\nu_1=135.4916\,\mathrm{d}^{-1}$, $\nu_2=135.7454\,\mathrm{d}^{-1}$, and a quintuplet of pulsation frequencies centered at $\nu_3=135.9233\,\mathrm{d}^{-1}$. The peak of the second pair of rotational sidelobes associated with $\nu_3$ is lower than 4.6 times the background noise level (but remains above 4 times the background noise level). Given their frequency spacing from the first pair of rotational sidelobes closely approximates the rotational frequency, we have incorporated them into our frequency detection results. The results for pulsation frequencies are presented in Table~\ref{Tab:2}. It is worth noting that the frequency separation $\nu_2-\nu_1=0.2538\,\mathrm{d}^{-1}$ is very close to the rotation frequency $\nu_{\mathrm{rot}}=0.25551\,\mathrm{d}^{-1}$. This suggests that $\nu_1$ and $\nu_2$ may be part of the frequency triplet of a dipole mode. After pre-whitening the pulsation frequencies, a slight power excess is evident in the residual amplitude spectrum, as shown in Fig.~\ref{Figure:2}~(d). The power excess indicates the presence of unresolved signals remaining in the amplitude spectrum. 

\subsubsection{TIC~318007796}\label{sect:PulFreAnalysis7796}

After high-pass filtering, the amplitude spectrum reveals the existence of two pulsation frequencies for TIC~318007796: a triplet centered around a frequency $\nu_1=192.73995\,\mathrm{d}^{-1}$ and a quintuplet centered at a higher frequency $\nu_2=196.33065\,\mathrm{d}^{-1}$, as depicted in Fig.~\ref{Figure:2}~(c). The detailed results of the pulsation frequencies are presented in Table~\ref{Tab:2}. Within the uncertain range, our results are consistent with those reported by \citet{2024MNRAS+527+9548H}. Upon pre-whitening the pulsation frequencies, a slight power excess is observed in the residual amplitude spectrum within the frequency ranges of $192.0-192.5\,\mathrm{d}^{-1}$ and $195.0-196.0\,\mathrm{d}^{-1}$, as shown in Fig.~\ref{Figure:2}~(d). These power excess indicate the presence of unresolved signals at these locations.

It is noteworthy that our investigation of TIC~96315731, TIC~72392575, and TIC~318007796 has exclusively unveiled p-mode pulsations indicative of the stellar outer layers, with no detection of g-modes sensitive to the core region \citep{2010aste+book+A}. Consequently, the task of ascertaining the complete rotational profile, spanning from the surface to the core, of these particular roAp stars remains a challenge within the scope of current asteroseismic analyses.

\section{Pulsation Mode and Geometry}\label{PulMGeo}

For roAp stars, the frequency triplet is commonly associated with the dipole mode, while the frequency quintuplet is typically linked to the quadrupole mode. However, some roAp stars, such as HD~6532 \citep{1996MNRAS+281+883K} and HR~3831 \citep{1990MNRAS+247+558K}, respectively exhibit the dipole mode as the pulsation modes of the frequency quintuplet and frequency septuplet. This is due to the influence of the magnetic field, which can increase the degeneracy of the pulsation mode, allowing dipole mode to exhibit five or more fine structures \citep{2004MNRAS+350+485S}. Therefore, solely relying on the number of rotational sidelobes in the pulsation frequency does not allow us to determine the pulsation mode of roAp stars. In this paper, we assume that the pulsation mode associated with the quintuplet of pulsation frequencies is a quadrupole mode. According to the oblique pulsator model, the rotational sidelobes of pure dipole or quadrupole mode are precisely split by the rotation frequency, with equal phase for all components of the multiplet at the maximum pulsation amplitude. To test this, we employ the rotation frequency fixed rotational sidelobes corresponding to the pulsation frequency. It is ensured that the interval between these rotational sidelobes precisely equates to the rotation frequency. Subsequently, a suitable time zero-point is chosen such that the phases of the first pair of rotational sidelobes for the pulsation frequency are equal, i.e., $\phi\left(\nu-\nu_{\mathrm{rot}}\right)=\phi\left(\nu+\nu_{\mathrm{rot}}\right)$. Finally, linear least squares fitting is employed to the combined light curves to obtain phase values for all components. The results of the linear least squares fitting for TIC~96315731, TIC~72392575, and TIC~318007796 are presented in Table~\ref{Tab:3}.

According to the oblique pulsator model, there are the following relationships between the amplitude ratio of the quintuplet of pulsation frequency and the stellar rotation inclination $\left(i\right)$ and magnetic obliquity angle $\left(\beta\right)$ for the quadrupole mode \citep{1990MNRAS+247+558K}:
\begin{equation}\label{eq:1}
  \frac{A_{+2}^{(2)}+A_{-2}^{(2)}}{A_{0}^{(2)}}=
  \frac{3\sin ^2\beta \sin ^2i}{(3\cos ^2\beta -1)(3\cos ^2i-1)},
\end{equation}
\begin{equation}\label{eq:2}
  \frac{A_{+1}^{(2)}+A_{-1}^{(2)}}{A_{0}^{(2)}}=
  \frac{12\sin \beta \sin i\cos \beta \cos i}{(3\cos ^2\beta -1)(3\cos ^2i-1)}.
\end{equation}

From Equations~\eqref{eq:1} and~\eqref{eq:2}, we can derive the standard constraints for the oblique pulsator:
\begin{equation}\label{eq:3}
  \tan i\tan \beta =4\frac{A_{+2}^{(2)}+A_{-2}^{(2)}}{A_{+1}^{(2)}+A_{-1}^{(2)}},
\end{equation}
where $A_{0}^{(2)}$, $A_{\pm 1}^{(2)}$, and $A_{\pm 2}^{(2)}$ represent the amplitudes of the central peak and the first and second pair of rotational sidelobes for the frequency quintuplet, respectively. The $\tan i\tan \beta$ provides constraints on the stellar rotation inclination $\left(i\right)$ and magnetic obliquity angle $\left(\beta\right)$, but these constraints are limited. We cannot directly determine the values of $i$ and $\beta$ from these constraints. Simultaneously, solving Equations~\eqref{eq:1} and~\eqref{eq:2} allows us to obtain the values of $i$ and $\beta$. These provide the geometry of the pulsation mode.

\subsection{TIC~96315731}\label{PulMGeo5731}

For TIC~96315731, the phases of the frequency quintuplet are unequal, as shown in Table~\ref{Tab:3}. This indicates that the mode corresponding to the frequency quintuplet is distorted. Additionally, the amplitudes of each pair of rotational sidelobes are unequal. This is attributed to the effect of the Coriolis force \citep{2002A&A+391+235B}. By substituting the amplitudes from Table~\ref{Tab:3} into Equations~\eqref{eq:1},~\eqref{eq:2}, and~\eqref{eq:3}, we determine that $\tan i\tan \beta =1.63 \pm 0.14$. Here, the stellar rotation inclination angle is $i=68.52^{\circ}\pm 0.92^{\circ}$ and the magnetic obliquity angle is $\beta =32.62^{\circ}\pm 1.29^{\circ}$, or vice versa.

\subsection{TIC~72392575}\label{PulMGeo2575}

For TIC~72392575, the phases of the frequency quintuplet are not equal but very closed, as depicted in Table~\ref{Tab:3}. This indicates a slight distortion in the mode corresponding to the frequency quintuplet. Similar to TIC~96315731, based on the amplitudes of the quintuplet from Table~\ref{Tab:3} and Equations~\eqref{eq:1},~\eqref{eq:2}, and~\eqref{eq:3}, we determine that $\tan i\tan \beta=1.73\pm 0.30$. Here, the stellar rotation inclination angle is $i=78.08^{\circ}\pm 1.15^{\circ}$ and the magnetic obliquity angle is $\beta =20.07^{\circ}\pm 1.50^{\circ}$, or vice versa.

\begin{table}[htbp]
  \centering
  \begin{threeparttable}
    \caption[]{Linear Least Squares Fitting Results for Pulsation Frequencies of TIC~96315731, TIC~72392575, and TIC~318007796, with Frequency Spacing Equal to the Rotation Frequency
    }\label{Tab:3}
    \begin{tabular}{clccc}
      \hline\noalign{\smallskip}
      Star                           & Label                          & Frequency                      & Amplitude                     & Phase                        \\
                                     &                                & $\left(\mathrm{d}^{-1}\right)$ & $\left(\mathrm{mmag}\right) $ & $\left(\mathrm{rad}\right) $ \\
      \hline\noalign{\smallskip}
      \multirow{5}{*}{TIC~96315731}  & $\nu-2\nu _{\mathrm{rot}}$     & 164.96122                      & 0.061$\pm$0.007               & 1.86 $\pm$0.12               \\
                                     & $\nu- \nu _{\mathrm{rot}}$     & 165.11106                      & 0.160$\pm$0.007               & -1.16$\pm$0.05               \\
                                     & $\nu                     $     & 165.26090                      & 0.109$\pm$0.007               & -1.93$\pm$0.07               \\
                                     & $\nu+ \nu _{\mathrm{rot}}$     & 165.41074                      & 0.140$\pm$0.007               & -1.16$\pm$0.05               \\
                                     & $\nu+2\nu _{\mathrm{rot}}$     & 165.56058                      & 0.061$\pm$0.007               & 2.71$\pm$0.12                \\
      \noalign{\smallskip}\hline\noalign{\smallskip}
      \multirow{7}{*}{TIC~72392575}  & $\nu_{1}                     $ & 135.4915                       & 0.022$\pm$0.005               & -1.43 $\pm$ 0.23             \\
                                     & $\nu_{2}                     $ & 135.7448                       & 0.054$\pm$0.005               & -0.91 $\pm$ 0.09             \\

                                     & $\nu_{3}-2\nu _{\mathrm{rot}}$ & 135.4123                       & 0.024$\pm$0.005               & 2.37  $\pm$ 0.21             \\
                                     & $\nu_{3}- \nu _{\mathrm{rot}}$ & 135.6678                       & 0.052$\pm$0.005               & 2.42  $\pm$ 0.10             \\
                                     & $\nu_{3}                     $ & 135.9233                       & 0.191$\pm$0.005               & 2.60  $\pm$ 0.03             \\
                                     & $\nu_{3}+ \nu _{\mathrm{rot}}$ & 136.1788                       & 0.052$\pm$0.005               & 2.42  $\pm$ 0.10             \\
                                     & $\nu_{3}+2\nu _{\mathrm{rot}}$ & 136.4343                       & 0.021$\pm$0.005               & 2.38  $\pm$ 0.24             \\
      \noalign{\smallskip}\hline\noalign{\smallskip}
      \multirow{8}{*}{TIC~318007796} & $\nu_{1}- \nu _{\mathrm{rot}}$ & 192.49245                      & 0.108$\pm$0.007               & -1.91 $\pm$ 0.06             \\
                                     & $\nu_{1}                     $ & 192.73995                      & 0.049$\pm$0.007               & -0.95 $\pm$ 0.14             \\
                                     & $\nu_{1}+ \nu _{\mathrm{rot}}$ & 192.98745                      & 0.093$\pm$0.007               & -1.67 $\pm$ 0.07             \\

                                     & $\nu_{2}-2\nu _{\mathrm{rot}}$ & 195.83565                      & 0.062$\pm$0.007               & 2.65  $\pm$ 0.11             \\
                                     & $\nu_{2}- \nu _{\mathrm{rot}}$ & 196.08315                      & 0.083$\pm$0.007               & -1.81 $\pm$ 0.08             \\
                                     & $\nu_{2}                     $ & 196.33065                      & 0.310$\pm$0.007               & -3.07 $\pm$ 0.02             \\
                                     & $\nu_{2}+ \nu _{\mathrm{rot}}$ & 196.57815                      & 0.107$\pm$0.007               & -1.81 $\pm$ 0.06             \\
                                     & $\nu_{2}+2\nu _{\mathrm{rot}}$ & 196.82565                      & 0.111$\pm$0.007               & -2.78 $\pm$ 0.06             \\
      \noalign{\smallskip}\hline
    \end{tabular}{\footnotesize \textbf{Note.} The time zero-points for fitting to TIC~96315731, TIC~72392575, and TIC~318007796 are BJD~2,458,861.12908, BJD~2,459,797.32875, and BJD~2,459,213.62661, respectively. The table is organized with columns for star names, frequency labels, frequencies, amplitudes, and phases.}
  \end{threeparttable}
\end{table}

\subsection{TIC~318007796}\label{PulMGeo7796}

For TIC~318007796, the phases of the frequency quintuplet are unequal, as shown in Table~\ref{Tab:3}. This implies that the mode corresponding to the frequency quintuplet is distorted. Similar to TIC~96315731, according to the amplitudes of the quintuplet from Table~\ref{Tab:3} and Equations~\eqref{eq:1},~\eqref{eq:2}, and~\eqref{eq:3}, we derive that $\tan i\tan \beta =3.64\pm 0.28$. Here, the stellar rotation inclination angle is $i=81.43^{\circ}\pm 0.51^{\circ}$, and the magnetic obliquity angle is $\beta=28.77^{\circ}\pm 0.66^{\circ}$, or vice versa. Within the uncertainty range, our result of $\tan i\tan \beta$ is consistent with the finding of \citet{2024MNRAS+527+9548H}, where they reported $\tan i\tan \beta =3.51\pm 0.27$. 

\section{Pulsation Amplitude and Phase Modulation}\label{PulAmpandPhaMod}

According to the oblique pulsator model, the pulsation nodes for a pure quadrupole mode are located at a colatitude of $\pm 54.7^{\circ}$, while for a pure dipole mode, the pulsation nodes coincide with the equator. If $i+\beta >54.7^{\circ}$ (or $90^{\circ}$, for a pure dipole mode) the pulsation nodes will cross the line of sight, resulting in alternating visibility of two pulsation poles. This causes the pulsation amplitude to be zero, accompanied by a phase reversal of $\pi\mbox{-}\mathrm{rad}$. By analyzing the pulsation amplitude and phase modulation curves, we can determine whether there is distortion in the pulsation mode. In addition, the variations in the rotational light are governed by the abundance spots on the stellar surface, while the changes in pulsation amplitude are controlled by the angle between the pulsation axis and the line of sight. Therefore, the phase difference between the maximum values of the rotational light curve and the pulsation amplitude indicates the relative positioning of the abundance spots and the pulsation poles.

To test this, we divide the combined light curves into segments of specified length based on the pulsation period. Then, we apply linear least squares fitting to each segment to obtain the amplitude and phase values corresponding to the pulsation frequency. These amplitude and phase data are employed to construct pulsation and phase modulation curves as a function of the rotation phase.

\subsection{TIC~96315731}\label{PulAmpPhaModul5731}

\begin{figure}[htbp]
  \centering
  \includegraphics[width=1\textwidth, angle=0]{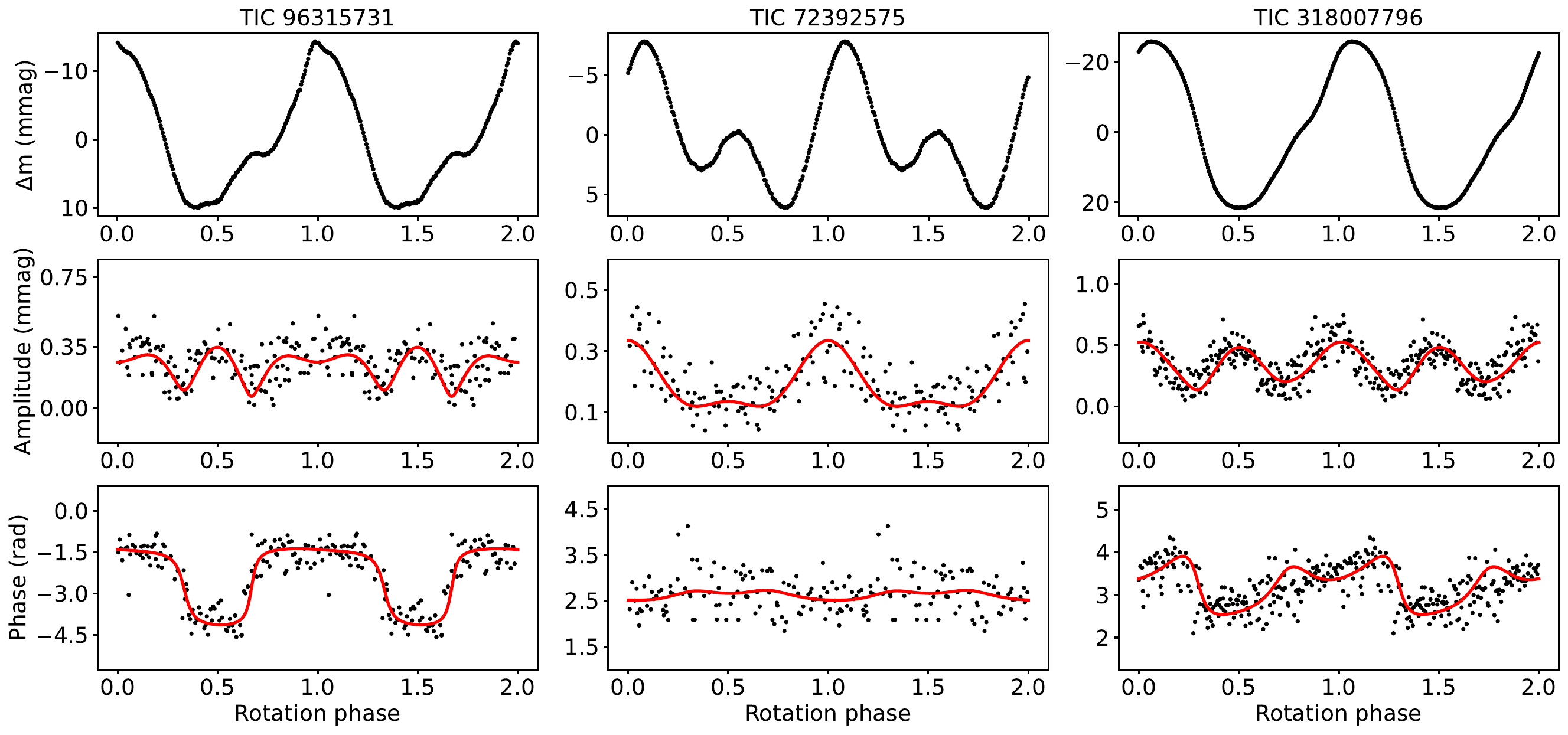}
  \caption{Phase folded light curves (top panel), the modulation curves of the pulsation amplitude (middle panel), and phase (bottom panel) as a function of rotation phase for TIC~96315731, TIC~72392575, and TIC~318007796. The bin size of the phase folded light curves is 0.005. Larger phase errors occur at smaller pulsation amplitudes, hence points with phase errors greater than $1\sigma$ have been excluded. The red curves represent the fitted curves of pulsation amplitude and phase modulation using the method proposed by \citet{1992MNRAS+259+701K}. The time zero-points for TIC~96315731, TIC~72392575, and TIC~318007796 are BJD~2,458,861.12908, BJD~2,459,797.32875, and BJD~2,459,213.62661, respectively.}\label{Figure:5}
\end{figure}

For TIC~96315731, we divide the combined light curve into 129 segments, with each segment having a length of 120 pulsation cycles, corresponding to a duration of 17.43 hours. Then, we apply linear least squares fitting to each segment, obtaining the amplitude and phase values corresponding to the pulsation frequency of $165.2609\,\mathrm{d}^{-1}$. Fig.~\ref{Figure:5} shows the rotational light curve, as well as the pulsation amplitude and phase modulation curves as a function of the rotation phase. At rotation phase 0, the phase of maximum pulsation amplitude closely approximates that of the maximum rotational light. This implies that the abundance spots are close to the pulsation poles (assuming the abundance spot near the magnetic poles). At this time, the angle between the pulsation axis and the line of sight is $i-\beta=35.90^{\circ}$. Following a half-cycle rotation, the secondary maximum of the pulsation amplitude occurs, with the angle between the pulsation axis and the line of sight being $i+\beta=101.14^{\circ}$. Furthermore, near the rotational phase 0.25, the pulsation amplitude reaches its minimum, accompanied by a phase reversal of $\pi\mbox{-}\mathrm{rad}$ in the pulsation phase. These characteristics are consistent with the expectations of the oblique pulsar model. For the quadrupole mode (or dipole mode), the minimum pulsation amplitude is non-zero, suggesting a distortion in the pulsation mode.

\subsection{TIC~72392575}\label{PulAmpPhaModul2575}

For TIC~72392575, we divide the combined light curve into 129 segments, each with a length of 100 pulsation cycles, corresponding to approximately 17.77 hours. Then, we perform a linear least squares fitting on each segment, obtaining the amplitude and phase values corresponding to the pulsation frequency of $135.9233\,\mathrm{d^{-1}}$. Fig.~\ref{Figure:5} displays the rotational light curve, as well as the modulation curve of the pulsation amplitude and phase as a function of rotation phase for the star. At rotation phase 0, the phase of the maximum pulsation amplitude is not consistent with the maximum of the rotation light, which indicates that the abundance spot is not aligned with the pulsation poles. This phenomenon is quite common in roAp stars, as evidenced by the studies on the roAp star HD~86181 by \citet{2021MNRAS+506+5629S} and on HD~12098 by \citet{2024MNRAS+529+556K}. At this moment, the angle between the pulsation axis and the line of sight, $i-\beta=58.01^{\circ}$, with the upper limb of the pulsation node appearing on the line of sight. At rotation phase 0.25, the pulsation amplitude reaches its minimum (which is not zero), with a slight variation in the pulsation phase but not a phase reversal of $\pi\mbox{-}\mathrm{rad}$. This implies that the quadrupole mode is distorted. At rotation phase 0.5, the pulsation amplitude reaches its second maximum, with the angle between the pulsation axis and the line of sight being $i+\beta=98.15^{\circ}$. The pulsation phase between 0.25 and 0.75 in the pulsation phase modulation curve appears to be more scattered compared to other positions. This phenomenon can be attributed to the inverse relationship between the uncertainty of pulsation phase and pulsation amplitude \citep{1999DSSN+13+28M}. We note that the pulsation amplitude and phase modulation curves of the star closely resemble those of two other roAp stars, KIC 10685175 \citep{2020ApJ+901+15S} and HD~86181 \citep{2021MNRAS+506+5629S}, pulsating in distorted quadrupole modes. This implies that our assumption may be correct.

\subsection{TIC~318007796}\label{PulAmpPhaModul7796}

Similar to the previous two stars, we divide the combined light curve of TIC~318007796 into 198 segments, each with a length of 50 pulsation periods, corresponding to a duration of 6.11 hours for each segment. We then apply linear least squares fitting to each segment at the pulsation frequency of $196.33065\,\mathrm{d^{-1}}$ to obtain the corresponding amplitude and phase values. The rotational light curve, along with the pulsation amplitude and phase modulation curves are shown in Fig.~\ref{Figure:5}. At rotation phase 0, the approximate equal phase between the maximum pulsation amplitude and the maximum of the rotating light indicates that the abundance spots are close to the pulsation poles. In this case, the angle between the pulsation axis and the line of sight, $i-\beta=52.66^{\circ}$. Around rotation phase 0.25, the pulsation amplitude reaches its minimum (which is not zero). Additionally, there is a variation in the pulsation phase, but it is less than $\pi\mbox{-}\mathrm{rad}$. This phase change indicates that the pulsation mode is distorted. After 0.25 rotation phase, the pulsation amplitude reaches its second maximum, which is approximately equal to the maximum pulsation amplitude. The angle between the pulsation axis and the line of sight becomes $i+\beta=110.20^{\circ}$. 

\section{Spherical Harmonic Decomposition}\label{SHD}

To further determine the contribution of the amplitude of pulsation modes associate with different spherical harmonic components ($l=$ 0, 1, 2, 3) to the amplitude spectra of these stars, we employ the axisymmetric spherical harmonic decomposition method for roAp star pulsation mode proposed by \citet{1992MNRAS+259+701K}. This method utilizes a linear combination of axisymmetric $\left( m=0 \right) l=$ 0, 1, 2, and 3 spherical harmonics to describe the amplitude spectrum of the pulsation mode. For the frequency quintuplet, it is sufficient to characterize the amplitude spectrum by employing a linear combination of spherical harmonic components with $l \leqslant 2$. The spherical harmonic decomposition results for the frequency quintuplets in TIC~96315731, TIC~72392575, and TIC~318007796 are presented in Table~\ref{Tab:4}. 

\begin{table}[htbp]
  \centering
  \begin{threeparttable}
    \caption[]{The Spherical Harmonic Decomposition Results of Frequency Quintuples for TIC~96315731, TIC~72392575, and TIC~318007796
    }\label{Tab:4}
    \begin{tabular}{cccccccc}
      \hline\noalign{\smallskip}
      Star                           & $l$ & $A_{-2}^{(l)}$     & $A_{-1}^{(l)}$    & $A_{0}^{(l)}$     & $A_{+1}^{(l)}$    & $A_{+2}^{(l)}$    & $\phi$           \\
                                     &     & $(\mathrm{mmag}) $ & $(\mathrm{mmag})$ & $(\mathrm{mmag})$ & $(\mathrm{mmag})$ & $(\mathrm{mmag})$ & $(\mathrm{rad})$ \\
      \hline\noalign{\smallskip}
      \multirow{3}{*}{TIC~96315731}  & 2   & 0.061              & 0.150             & -0.109            & 0.150             & 0.061             & 1.860            \\
                                     & 1   &                    & 0.309             & 0.368             & 0.289             &                   & -1.219           \\
                                     & 0   &                    &                   & 0.400             &                   &                   & 2.084            \\
      \noalign{\smallskip}\hline\noalign{\smallskip}
      \multirow{3}{*}{TIC~72392575}  & 2   & 0.024              & 0.054             & -0.191            & 0.05              & 0.021             & 2.37             \\
                                     & 1   &                    & 0.003             & 0.004             & 0.003             &                   & -1.737           \\
                                     & 0   &                    &                   & 0.381             &                   &                   & 2.477            \\
      \noalign{\smallskip}\hline\noalign{\smallskip}
      \multirow{3}{*}{TIC~318007796} & 2   & 0.062              & 0.082             & -0.310            & 0.108             & 0.111             & 2.650            \\
                                     & 1   &                    & 0.130             & 0.082             & 0.170             &                   & -1.158           \\
                                     & 0   &                    &                   & 0.647             &                   &                   & 2.828            \\
      \noalign{\smallskip}\hline
    \end{tabular}{\footnotesize \textbf{Note.} The time zero-points for TIC~96315731, TIC~72392575, and TIC~318007796 are BJD~2,458,861.12908, BJD~2,459,797.32875, and BJD~2,459,213.62661, respectively. The rotation inclination ($i$) and magnetic obliquity ($\beta$) angles are $i=68.52^\circ$ and $\beta=32.62^\circ$ for TIC~96315731, $i=78.08^\circ$ and $\beta=20.07^\circ$ for TIC~72392575, and $i=81.43^\circ$ and $\beta=28.77^\circ$ for TIC~318007796. The table is organized into columns listing star names, spherical harmonic degrees $(l)$ for different pulsation modes, amplitudes for azimuthal orders ranging from $m=-2$ to $m=+2$, and phases for each spherical harmonic degree.}
  \end{threeparttable}
\end{table}

\subsection{TIC~96315731}\label{SHD5731}

Our results for TIC~96315731 indicate that the dipole $\left( l=1 \right)$ mode contributes 0.966 mmag to the pulsation amplitude. Additionally, the radial $\left( l=0 \right)$ mode contributes 0.400 mmag, while the quadrupole $\left( l=2 \right)$ mode contributes 0.313 mmag. Among all the modes, the dipole mode dominates the contribution of the amplitude, indicating that the pulsation mode of the star is a distorted dipole mode rather than a distorted quadrupole mode. It is worth noting that our result differs from that reported by \citet{2021MNRAS+506+1073H} for TIC~96315731. In their work, they leaned towards a distorted quadrupole mode based on the analysis of pulsation amplitudes and phases modulation curves. According to the spherical harmonic decomposition results presented in Table~\ref{Tab:4}, we model the pulsation amplitude and phase modulation curve for TIC~96315731, illustrating by the red curves in Fig.~\ref{Figure:5}. 

\subsection{TIC~72392575}\label{SHD2575}

For TIC~72392575, our results reveal that the contributions of the radial $\left( l=0 \right)$ mode, dipole $\left( l=1 \right)$ mode, and quadrupole $\left( l=2 \right)$ mode to the pulsation amplitude are 0.381 mmag, 0.01 mmag, and -0.042 mmag, respectively, at the maximum of pulsation amplitude. The radial mode dominates contribution of the amplitude, followed by the quadrupole and dipole modes. Because the pulsation mode of the roAp stars are non-radial, the contribution of the dipole mode to the amplitude can be neglected compared to the quadrupole mode. Therefore, our hypothesis is valid, indicating that the pulsation mode of the star is a distorted quadrupole mode. We model the pulsation amplitude and phase modulation curves for the star using the spherical harmonic decomposition results presented in Table~\ref{Tab:4}. The results are shown by the red curves in Fig.~\ref{Figure:5}.

\subsection{TIC~318007796}\label{SHD7796}

According to the spherical harmonic decomposition results, at the maximum pulsation amplitude, the radial $\left( l=0 \right)$ mode, dipole $\left( l=1 \right)$ mode, and quadrupole $\left( l=2 \right)$ mode contribute 0.647 mmag, 0.382 mmag, and 0.053 mmag to the pulsation amplitude, respectively. After excluding the radial mode, the contribution of the dipole mode to the pulsation amplitude is greater compared to the quadrupole mode. Therefore, we conclude that the pulsation mode corresponding to the frequency quintuplet is a distorted dipole mode. Utilizing the spherical harmonic decomposition results from Table~\ref{Tab:4}, we model the pulsation amplitude and phase modulation curves for the star, which are shown in Fig.~\ref{Figure:5} as red curves.

\section{Conclusions}\label{DisandConclus}

TIC~96315731 is an roAp star with a quintuplet of pulsation frequencies. We determine its rotation frequency to be $\nu_{\mathrm{rot}}=0.1498360\pm 0.0000005\,\mathrm{d}^{-1}$ $\left( P_{\mathrm{rot}}=6.67396\pm 0.00002\,\mathrm{d} \right)$ and its pulsation frequency to be $\nu =165.2609\,\mathrm{d}^{-1}$ through careful analysis of the combined light curve. Then, we determine the rotation inclination angle $\left(i = 68.52^{\circ}\right)$ and the magnetic obliquity angle $\left(\beta = 32.62^{\circ}\right)$ through the amplitude ratio of the quintuplet of pulsation frequencies, which provides initial constraints on the stellar geometry.

TIC~72392575 is also an roAp star with a quintuplet of pulsation frequencies. We determine its rotation frequency to be $\nu_{\mathrm{rot}}=0.25551\pm 0.00002\,\mathrm{d}^{-1}$ $\left(P_{\mathrm{rot}}=3.9138\pm 0.0004\,\mathrm{d}\right)$. The pulsation frequencies of the star display intriguing characteristics. In the second pair of rotational sidelobes associated with the pulsation frequency $\nu_3=135.9233\,\mathrm{d}^{-1}$, two frequencies are observed, with a separation approximately equal to the rotation frequency. These signals may correspond to the frequencies originating from a dipole mode. According to the amplitude ratio of the quintuplet of pulsation frequencies, we determine the rotational inclination angle $\left(i=78.08^{\circ}\right)$ and the magnetic obliquity angle $\left(\beta=20.07^{\circ}\right)$.

TIC~318007796 is an roAp star with two pulsation modes. We derive its rotation frequency as $\nu_{\mathrm{rot}}=0.2475021\pm 0.0000004\,\mathrm{d}^{-1}$ $\left(P_{\mathrm{rot}}=4.040370\pm 0.000007\,\mathrm{d}\right)$, and the central peaks of the triplet and quintuplet of pulsation frequencies corresponded to frequencies $\nu_1=192.73995\,\mathrm{d}^{-1}$ and $\nu_2=196.33065\,\mathrm{d}^{-1}$, respectively. From the amplitude ratio of the quintuplet of the pulsation frequencies $\nu_2$, we obtain the rotation inclination angle $\left(i=81.43^{\circ}\right)$ and the magnetic obliquity angle $\left(\beta=28.77^{\circ}\right)$. 

We examine the phases of the frequency quintuplets from the linear least squares fitting results, along with the pulsation amplitude and phase modulation curves of TIC~96315731, TIC~72392575, and TIC~318007796. The results suggest the presence of distortion in the pulsation modes.Integrating these findings with the results of spherical harmonic decomposition, we confirm that the pulsation modes of the frequency quintuplets corresponding to TIC~96315731, TIC~72392575, and TIC~318007796 are distorted dipole modes, distorted quadrupole modes, and distorted dipole modes, respectively.

\begin{acknowledgements}
This work includes data collected by the TESS mission. Funding for the TESS mission is provided by the NASA Explorer Program. We thank Shi Fangfei for the valuable assistance provided during the data processing phase. This work received the generous support of the National Natural Science Foundation of China under grants Nos. U2031204, 12373038, 12163005, and 12288102, the science research grants from the China Manned Space Project with No. CMSCSST-2021-A10, the Natural Science Foundation of Xinjiang Nos. 2022D01D85 and 2022TSYCLJ0006.
\end{acknowledgements}

\bibliographystyle{raa}
\bibliography{bibtex}
\end{document}